\documentclass[12pt]{article}
\pdfoutput = 1
\usepackage{hyperref}
\usepackage{graphicx} 
\begin{document}
\title{Accretion onto a noncommutative inspired Schwarzschild black hole}
\author{
{\bf {\normalsize Biplab Paik}$^{a}
$\thanks{withbiplab@gmail.com}},
{\bf {\normalsize Sunandan Gangopadhyay}
$^{b,c}$\thanks{sunandan.gangopadhyay@gmail.com, sunandan@iiserkol.ac.in, sunandan@associates.iucaa.in}},\\[0.2cm]
$^{a}$ {\normalsize Rautara MNM High School, Habra, North 24 PGS, West Bengal, India}\\[0.2cm]
$^{b}$ {\normalsize Department of Physical Sciences,}\\
{\normalsize Indian Institute of Science Education and Research Kolkata} \\
{\normalsize Mohanpur 741246, Nadia, West Bengal, India}\\[0.2cm]
$^{c}${\normalsize Visiting Associate in Inter University Centre for Astronomy $\&$ Astrophysics (IUCAA),}\\
{\normalsize Pune 411007, India}\\[0.3cm]
}
\date{}

\maketitle

\begin{abstract}
\noindent In this paper we investigate the problem of ordinary baryonic matter accretion onto the noncommutative geometry inspired Schwarzschild black hole. The fundamental equations governing the spherically symmetric steady state matter accretion are deduced. These equations are seen to be modified due to the presence of noncommutativity. The matter accretion rate is computed and is found to increase rapidly with the increase in strength of the noncommutative parameter.  The sonic radius reduces while the sound speed at the sonic point increases with the increase in the strength of noncommutativity. The profile of the thermal environment is finally investigated below the sonic radius and at the event horizon and is found to be affected by noncommutativity. 
\end{abstract}

\section{Introduction} In the $20^{th}$ century it was first realized that it is not nuclear fusion but actually gravity which  powers most of the luminous objects in our Universe. The mechanism by which gravity does this is the accretion of matter onto gravitating bodies. Of course stronger gravitating bodies are bound to be more effective in releasing more gravitational power. It is thus obvious that the accretion of matter onto black holes, which are known to be the strongest gravitating objects in the universe, is responsible for the high-energy flux from active galactic nuclei and quasars. The investigations of the accretion processes onto celestial objects were initiated by Hoyle and Lyttleton in 1939 $\cite{Hoyle}$, later studies were carried out by Bondi and  Hoyle $\cite{Bondi_Hoyle}$ for a pressure-less gas falling onto a moving star. Subsequently, a theory of stationary, spherically symmetric and transonic hydrodynamic accretion of adiabatic fluid onto a gravitating astrophysical body at rest was formulated in a seminal paper by Bondi in 1952 $\cite{BNDI}$, in a purely Newtonian way. This calculation included the effect of pressure of the in-falling material. Thereafter a large body of literature has been devoted to the theoretical and observational studies of accretion processes (see for example the standard texts $\cite{SHP_TEU}, \cite{Frank}$). The general relativistic treatment of the problem of steady-state spherically symmetric flow of a test gas onto a Schwarzschild black hole was initiated in $\cite{MCHL}$. There the equations of motion for steady-state spherically symmetric flow of matter into (or out) of a black hole had been solved for simple polytropic gases. It was argued that infalling matter would be heated to temperatures as hot as $10^{12}$K and the resulting  X-ray luminosities were shown to be of the order of $10^{37}$ $erg/sec$. Further, comprehensive analysis has also been carried out to compute the luminosity and frequency spectrum $\cite{SHPRO1}$ and study the influence of an interstellar magnetic field on the accretion of ionized gases $\cite{SHPRO2}$. Aspects of matter accretion onto a charged black hole that was originally considered in $\cite{MCHL}$, was more elaborately investigated in $\cite{Freitas}$. Studies involving spherical winds and shock transitions were carried out in $\cite{Mathews}$. Efforts to understand the influence of back-reaction on the phenomenon of accretion can be found in \cite{Malec}-\cite{Babichev}. Moreover, in recent times, accretion of a hypothetical phantom fluid onto black holes has been investigated in \cite{Babichev2}-\cite{Gao}.

 The increasing recognition of the importance of accretion has accompanied the dramatic expansion of observational techniques in astronomy, in particular it helps in explaining the full range of the electromagnetic spectrum from the radio to X-rays and $\gamma$-rays. It is worth emphasizing that not only in the large length scale but even in the Planck scale the phenomenon of accretion may lead to provide important insights in testing the physical principles. Indeed, aspects of mini black holes are becoming an important topic of research, since it is speculated that they may be produced at the large hadron collider. If this really happens then even a shortly existing thermal environment around such mini black holes may lead to some new features in  the process of accretion. A few initiatives had actually been taken in this direction, namely, the investigation of accretion phenomenon in higher dimensional spherically symmetric black holes $\cite{John}$ and in a background string cloud model $\cite{Maharaj}$. It was shown in $\cite{John}$ that the accretion rate decreases with increase in the dimension of spacetime. In the background string cloud model $\cite{Maharaj}$, it was observed that the matter accretion rate increases with the increasing string cloud parameter. In $\cite{Yang}$ quantum gravity corrections has been included in the process of matter accretion onto a Schwarzschild black hole. 
 
 Noncommutative geometry $\cite{Alain}$ has been a very active area of research because of its direct relevance in Planck scale physics. The idea of noncommutative spacetime
 \begin{equation}
 [x_{\mu},x_{\nu}]=i\vartheta_{\mu \nu}
 \label{noncomm_rln}
 \end{equation} where $\vartheta_{\mu \nu}$ is an antisymmetric matrix representing the fundamental cell discretization of spacetime, was first put forward formally in $\cite{Snyder}$ but was largely ignored. It gained importance from string theory which pointed out that target spacetime coordinates become noncommuting operators on D-brane $\cite{D-brane}$. In this paper we study the phenomenon of accretion around a noncommutative inspired Schwarzschild black hole $\cite{Nicolini, SG}$. This black hole metric was first put forward in $\cite{Nicolini}$. It resulted from the argument that point-like structures get replaced by smeared objects due to noncommutativity $\cite{Spallucci, Spallucci2}$. The effect of smearing is to use a Gaussian distribution of minimal width $\sqrt{\vartheta}$ in place of a position Dirac-delta function. This observation resulted in choosing the mass density of a static, spherically symmetric, smeared, particle-like gravitational source as
 \begin{equation}
 \rho_{\vartheta}=\frac{M}{(4\pi \vartheta)^{3/2}}e^{-r^2/4\vartheta}.
 \label{density}
 \end{equation} A justification of choosing the above form for the mass density was provided in $\cite{SG}$. There the importance of the Voros product in defining the mass density was elaborated. In this paper our aim is to find the effect of the noncommutative parameter $\vartheta$ on the matter accretion rate. We further investigate the effect of noncommutativity on the sonic radius, the speed of sound and also the thermodynamic profiles of the accreting gas, namely, the gas density and the gas temperature.

The paper is organized as follows. In the following section we deduce the general equations for spherically symmetric accretion of matter onto a noncommutative inspired Schwarzschild black hole. In section 3, the sonic point is determined for this metric. In section 4, we obtain a polytropic solution to the accretion rate of matter and also carry out numerical calculations. The thermodynamic profiles of the accreting fluid is analyzed in the sub-Bondi regime in section 5 and at the event horizon in section 6. We conclude in section 7.

\section{General equations for spherical accretion} We now present the basic equations describing the phenomenon of spherical accretion to investigate the flow of gas onto the noncommutative(NC) inspired Schwarzschild black hole. The noncommutative inspired Schwarzschild black hole metric is given by $\cite{Nicolini}$
\begin{equation}
ds^2=-\left(1-\frac{2m(r)}{r}\right)dt^2+\left(1-\frac{2m(r)}{r}\right)^{-1}dr^2+r^2d\Omega^2
\label{Metric}
\end{equation}\\
where \[m(r)=\frac{2M}{\sqrt{\pi}}\gamma(3/2,r^2/4\vartheta)\] 
and $\gamma$ is the incomplete gamma function \[\gamma(s,x)\;=\; \int ^x_0 p^{s-1}e^{-p}dp.\] Our aim is to find how the noncommutativity of spacetime affects the matter accretion rate $\dot{M}$ onto the NC inspired Schwarzschild black hole, the asymptotic compression ratio and the temperature profiles below the sonic radius and at the event horizon. We follow the approach developed in $\cite{MCHL}$ to tackle the problem of relativistic spherical accretion. The accreting gas is considered to be a perfect fluid described by the energy momentum tensor
\begin{equation}
T^{\mu \nu}=(\rho+p)u^{\mu}u^{\nu}+pg^{\mu\nu}
\label{Perfect_fluid}
\end{equation} where $\rho$ and $p$ are the proper fluid energy density and pressure respectively, and 
\begin{equation}
u^{\mu}=\frac{dx^{\mu}}{ds}
\end{equation} is the fluid 4-velocity obeying the normalization condition $u^{\mu}u_{\mu}=-1$. \\

\noindent
The phenomenon of accretion is based on two important conservation laws. Considering that particle number is conserved, we have
 \begin{equation}
\nabla_{\mu}J^{\mu}=\nabla_{\mu}(n u^{\mu})=\:0
\label{Cons1}
\end{equation} where $n$ is the proper baryon number density and $J^{\mu}=nu^{\mu}$ is the baryon flux density, both the quantities being measured in the local inertial rest frame of the fluid. \\

\noindent
The conservation law of energy-momentum reads 
\begin{equation}
{\nabla_{\mu}T^{\mu}}_{\nu}=0
\label{Cons2}
\end{equation} The nonzero components of particle 4-velocity are $u^0=(dt/ds)$ and $u^1= v=dr/ds$. Now using the relation $u^{\mu}u_{\mu}=-1$, one gets for a NC inspired Schwarzschild black hole
\begin{equation}
u^0=\left\lbrace  \frac{v^2+(1-2m(r)/r)}{(1-2m(r)/r)^2} \right\rbrace^{1/2}.
\end{equation} For steady state spherical accretion, eq.($\ref{Cons1}$) can be written as
\begin{equation}
\frac{1}{\sqrt{g}}\frac{\partial}{\partial r}(\sqrt{g}J^{r})=0 
\label{Cons1x}
\end{equation} while the $\nu=0$ component of eq.($\ref{Cons2}$) gives the energy flux conservation
 \begin{equation}
\frac{1}{\sqrt{g}}\frac{\partial}{\partial r}(\sqrt{g}{T^{r}}_{0})=0.
\label{Cons2x}
\end{equation} Noting that $J^r=nu^r$ and ${T^r}_{0}=(\rho+p)g_{00}u^ru^0$, eq.(s)($\ref{Cons1x}$, $\ref{Cons2x}$) take the form
\begin{equation}
\frac{1}{r^2} \frac{\partial}{\partial r} \left\lbrace r^2 n u^r\right\rbrace=0
\label{10}
\end{equation} 
\begin{equation}
\frac{1}{r^2} \frac{\partial}{\partial r} \left\lbrace r^2(\rho+p)u^r\left(1-\frac{2m(r)}{r}+(u^r)^2  \right)\right\rbrace=0.
\label{11}
\end{equation} Integrating eq.(s)($\ref{10}$, $\ref{11}$) yields
\begin{equation}
r^2 nv=C_1
\label{Cons1y}
\end{equation}
\begin{equation}
r^2(\rho+p)v\left( 1-\frac{2m(r)}{r}+v^2  \right)^{1/2}=C_2
\label{Cons2y}
\end{equation} where $C_1$ and $C_2$ are constants of integration. Dividing eq.($\ref{Cons2y}$) by eq.($\ref{Cons1y}$) and squaring the expression gives  
\begin{equation}
\left( \frac{p+\rho}{n} \right)^2\;\left(1-\frac{2m(r)}{r}+v^2  \right)\quad=C_3
\label{Cons_G1}
\end{equation} where $C_3=(C_2/C_1)^2=constant$ . The above equation is known as the Bernoulli's equation for gas-flow. At $r=\infty$, $v=0$ and hence 
\begin{equation}
C_3=\left( \frac{\rho_{\infty}+p_{\infty}}{n_{\infty}}  \right)^2.
\label{15}
\end{equation} This yields
\begin{equation}
\;\left( \frac{p+\rho}{n} \right)^2\;\left(1-\frac{2m(r)}{r}+v^2  \right)=\left( \frac{p_{\infty}+\rho_{\infty}}{n_{\infty}} \right)^2.
\label{Cons_G}
\end{equation}
\section{Accretion onto noncommutative inspired black hole} The analysis of mass accreting into a spherically symmetric black hole starts by taking the accreting fluid to be adiabatic. Since there is no entropy production for an adiabatic fluid, the conservation of mass-energy is governed by the thermodynamic equation
 \begin{equation}
Tds=0=d\left(\frac{\rho}{n} \right)+pd\left( \frac{1}{n} \right).
\label{ADIAB}
\end{equation} This may easily be put in the form
\begin{equation}
\frac{d\rho}{dn}=\frac{\rho+p}{n}~.
\end{equation} The adiabatic sound speed `$a$' is defined as $\cite{SHP_TEU}$ 
 \begin{equation}
a^2\equiv \frac{dp}{d\rho}=\frac{dp}{dn}\frac{n}{p+\rho}~.
\label{S_speed}
\end{equation} We now take a differential of eq.($\ref{Cons1y}$) to obtain
 \begin{equation}
\frac{1}{n}dn+\frac{1}{v}dv+\frac{2}{r}dr=0.
\label{D_Cons}
\end{equation}  
For a perfect fluid ($\ref{Perfect_fluid}$), in the background of the NC inspired Schwarzschild black hole, the $\nu=1$ component of the energy-momentum conservation law ($\ref{Cons2}$) can be written as
\begin{equation}
v\frac{dv}{dr}\:=\:-\frac{dp}{dr}\,\left\lbrace \frac{1-2m(r)/r+v^2}{\rho+p}  \right\rbrace \,+ \,\frac{d}{dr}\left\lbrace \frac{m(r)}{r}\right\rbrace,
\end{equation} that is
\begin{equation}
v\frac{dv}{dr}=-\frac{dp}{dr}\left\lbrace \frac{1-2m(r)/r+v^2}{\rho+p}  \right\rbrace -\frac{m(r)}{r^2}+\frac{m'(r)}{r}~.
\end{equation} This can further be recast as 
\begin{equation}
vv'+a^2\left\lbrace1-2m(r)/r+v^2\right\rbrace \frac{n'}{n}=-\frac{m(r)}{r^2}+\frac{m'(r)}{r} 
\label{v_eqn}
\end{equation} where $v'=dv/dr$ , $n'=dn/dr$ and $m'=dm/dr$. Note that for the ordinary Schwarzschild black hole, $m(r)=M$. This then leads to the standard form for eq.($\ref{v_eqn}$), which is 
\begin{equation}
vv'+a^2\left\lbrace1-2M/r+v^2\right\rbrace \frac{n'}{n}=-\frac{M}{r^2}~.
\label{v_eqnSw}
\end{equation} It is evident that eq.($\ref{v_eqn}$) is different from eq.($\ref{v_eqnSw}$) due to the presence of the extra term $m'(r)$ and also the functional dependence on $r$ of $m$. The analysis involves investigating the consequential effects upon the matter accretion phenomenon.\\

\noindent Solving eq.(s) ($\ref{D_Cons}$) and ($\ref{v_eqn}$) yields 
\begin{equation}
v' = \frac{N_1}{N}~;~ n'=-\frac{N_2}{N}
\label{vn_primes}
\end{equation} where\begin{equation}
N_1= \frac{1}{n}\left[ \frac{2a^2}{r}\left\lbrace 1-\frac{2m(r)}{r}+v^2 \right\rbrace +  \frac{d}{dr}\left(\frac{m(r)}{r} \right)  \right]
\label{N_1}
\end{equation}
\begin{equation}
N_2= \frac{1}{v}\left[ \frac{d}{dr}\left(-\frac{m(r)}{r} \right) -\frac{2v^2}{r} \right]\;
\label{N_2}
\end{equation}
\begin{equation}
N= \frac{v^2 -a^2 \left\lbrace 1-\frac{2m(r)}{r}+v^2 \right\rbrace}{vn}~.
\label{N}
\end{equation}
Demanding the gas-flow to be subsonic $(v<a)$ at large $r$, there the subluminal $(a<1)$ speed of sound guarantees $v^2<<1$. In the large $r$ regime, the denominator in eq.($\ref{vn_primes}$) therefore becomes 
\begin{equation}
N\approx \frac{v^2-a^2}{v n} ~.
\end{equation} Hence $N<0$ for $r\rightarrow \infty$. On the other hand at the NC event horizon, $r_H=2m(r_H)$, it can be observed that 
\begin{equation}
N=\frac{v^2(1-a^2)}{vn}~.
\end{equation} 
Therefore in the near horizon regime the causality constraint $a^2 < 1$ provides us $N >0$. These characteristics point out that we must have $N =0$ for some critical radius $r=r_s$, $r_H < r_s < \infty$.  This essentially implies that the flow must pass through a critical point outside the event horizon. To have smoothness in the flow of the accreting fluid, we must have $N_1=N_2=N=0$ at $r=r_s$, that is
\begin{equation}
\frac{1}{n_s}\left[ \frac{2a_s^2}{r_s}\left\lbrace 1-\frac{2m(r_s)}{r_s}+v_s^2 \right\rbrace +  \left\lbrace \frac{d}{dr}\left(\frac{m(r)}{r} \right) \right\rbrace_{r=r_s} \right]=0
\label{N_1=0}
\end{equation}
\begin{equation}
\frac{1}{v_s}\left[ \left\lbrace\frac{d}{dr}\left(-\frac{m(r)}{r} \right)\right\rbrace_{r=r_s} -\frac{2v_s^2}{r_s} \right]=0
\label{N_2=0}
\end{equation}
 \begin{equation}
\frac{v_s^2 -a_s^2 \left\lbrace 1-\frac{2m(r_s)}{r_s}+v_s^2 \right\rbrace}{v_sn_s}=0.
\label{N=0}
\end{equation}
At the critical point $r=r_s$, we have from eq.($\ref{N_1=0}$) 
\begin{equation}
v_s^2=\frac{r_s}{2}~\frac{d}{dr}\left\lbrace-\frac{m(r)}{r}\right\rbrace_{r=r_s} ~.
\label{DMY_vs}
\end{equation} The above equation can be simplified further and takes the form 
\begin{eqnarray}
v^2_s &=& \frac{m(r_s)}{2r_s}\left\lbrace  1-\frac{r_s\gamma'_s}{\gamma_s} \right\rbrace \nonumber\\
 &=& \frac{M\gamma(3/2,r_s^2/4\vartheta)}{\sqrt{\pi}r_s}\left\lbrace  1-\frac{r_s^3}{4\vartheta^{3/2}}\frac{e^{-r^2_s/4\vartheta}}{\gamma(3/2,r_s^2/4\vartheta)} \right\rbrace ~ .
\label{v_s1}
\end{eqnarray} 
where \begin{equation}
\gamma_s= \gamma(3/2,r^2_s/4\vartheta) ~~;~~ \gamma'_s= \left[\frac{d}{dr}\gamma(3/2,r^2/4\vartheta)\right]_{r=r_s}. 
\label{symbols}
\end{equation} 
 One may easily check that for $r_s>>\sqrt{\vartheta}$, one recovers the commutative result $v^2_s=M/2r_s$. From eq.($\ref{N=0}$), we may relate $a_s$ to $v_s$ as 
 \begin{equation}
a_s^2[1-2m(r_s)/r_s+v^2_s]=v^2_s ~.
\end{equation}
 Replacing $m(r_s)/r_s$ in terms of $v_s$ by using eq.($\ref{v_s1})$ yields 
 \begin{equation}
a_s^2=\frac{v^2_s}{1-v_s^2\left\lbrace \frac{3+r_s\gamma'_s/\gamma_s}{1-r_s\gamma'_s/\gamma_s} \right\rbrace}~.
\end{equation} 
Writing this relation with $v_s$ being expressed in terms of $a_s$ gives 
\begin{equation}
v^2_s=\frac{a^2_s}{1+a^2_s\left\lbrace \frac{3+r_s\gamma'_s/\gamma_s}{1-r_s\gamma'_s/\gamma_s} \right\rbrace}~.
\label{v_s2}
\end{equation}  Therefore in the limit $a_s^2<<1$ we have
\begin{equation}
v^2_s=a_s^2.
\end{equation} In the commutative limit ($\vartheta \rightarrow 0$) (that is $(r_s\gamma_s'/\gamma_s)\rightarrow 0$), we recover the standard result $\cite{SHP_TEU}$ 
\begin{equation}
v^2_s=\frac{a^2_s}{1+3a^2_s}~.
\label{comm_vs}
\end{equation}
Note that at the sonic point the speed of sound would be low enough to ensure $a_s^2<<1$. This leads to $v_s^2\approx a_s^2$.\\

\noindent
The sonic radius in the noncommutative case can be obtained by equating eq.($\ref{v_s1}$) with eq.($\ref{v_s2}$), which yields
\begin{equation}
r_s=\frac{m(r_s)}{2}\frac{\left[1+3a^2_s-r_s\gamma'_s/\gamma_s(1-a^2_s)\right]}{a^2_s} ~.
\label{r_s1}
\end{equation} It can be readily seen that we have a transcendental equation for $r_s$ . However, for $a^2_s<<1$, which is indeed a standard approximation for investigating the accretion phenomenon, we obtain the following transcendental equation for $r_s$.
\begin{equation}
r_s= \frac{m(r_s)}{2a^2_s}[1-r_s\gamma'_s/\gamma_s].
\label{r_s2}
\end{equation}  The above expression implies that the effect of noncommutativity in spacetime leads to decrease in the sonic radius as compared to its commutative counterpart. In the $r_s>>\sqrt{\vartheta}$ limit, one gets back the usual result for accretion onto a Schwarzschild black hole,  $r_s= M/2a_s^2$. \\

\noindent
Now we are finally in a position to fully use the sonic conditions in order for computing the most relevant quantity namely the mass accretion rate. Assuming the average mass per gas particle to be $m_b$, eq.($\ref{Cons1y}$) provides the steady mass accretion rate to be 
\begin{equation}
\dot{M}=4\pi r^2 m_b n(r) v(r)=4\pi r^2_s m_bn(r_s)v(r_s) ~.
\label{RATE1}
\end{equation} 

\section{Mass accretion rate with polytropic equation of state} In order to calculate $\dot{M}$ explicitly eq.($\ref{Cons_G}$) and ($\ref{RATE1}$) must be supplemented with an equation of state. The equation of state that one introduces is a polytropic equation of state $\cite{BNDI}$
\begin{equation}
p=Kn^{\Gamma}
\label{p_n}
\end{equation} where $K$ and the adiabatic index $\Gamma$ are constants. Substituting this in eq.($\ref{ADIAB}$), we obtain \begin{equation}
\rho=\frac{K}{\Gamma-1}n^{\Gamma} +m_bn
\label{rho}
\end{equation} where $m_b$ is the integration constant. Note that $m_bn$ is the rest-mass energy density of the baryons eq.($\ref{S_speed}$) together  with the above expressions for $\rho$ and $p$ yields the following relation
 \begin{equation}
\Gamma K n^{\Gamma-1}=\frac{a^2m_b}{ 1- a^2/(\Gamma-1)}~.
\label{n_a}
\end{equation} Eq.($\ref{Cons_G}$) can be rewritten using eq.(s) ($\ref{rho}$), ($\ref{n_a}$)  as 
\begin{equation}
\left(1+\frac{a^2}{\Gamma-1-a^2}\right)^2\left(1-\frac{2m(r)}{r}+v^2  \right)=\left(1+\frac{a_{\infty}^2}{\Gamma-1-a_{\infty}^2}\right)^2.
\label{Dummy_Brnlli}
\end{equation}
 At the sonic radius $r=r_s$, the above equation can be put in the form (using the sonic velocity ($\ref{v_s2}$))
\begin{equation}
\left( 1\,+\,a^2_s\left\lbrace \frac{3+r_s\gamma'_s/\gamma_s}{1-r_s\gamma'_s/\gamma_s} \right\rbrace  \right)\;\left( 1-\frac{a^2_s}{\Gamma-1} \right)^2\;=\quad\left( 1-\frac{a^2_{\infty}}{\Gamma-1} \right)^2.
\label{BNLI}
\end{equation} The above equation is the the Bernoulli's equation for a NC inspired Schwarzschild black hole geometry. In the limit $r_s>> \sqrt{\vartheta}$,  the commutative version of Bernoulli's equation is recovered.\\

\begin{figure}
\includegraphics[scale=.15]{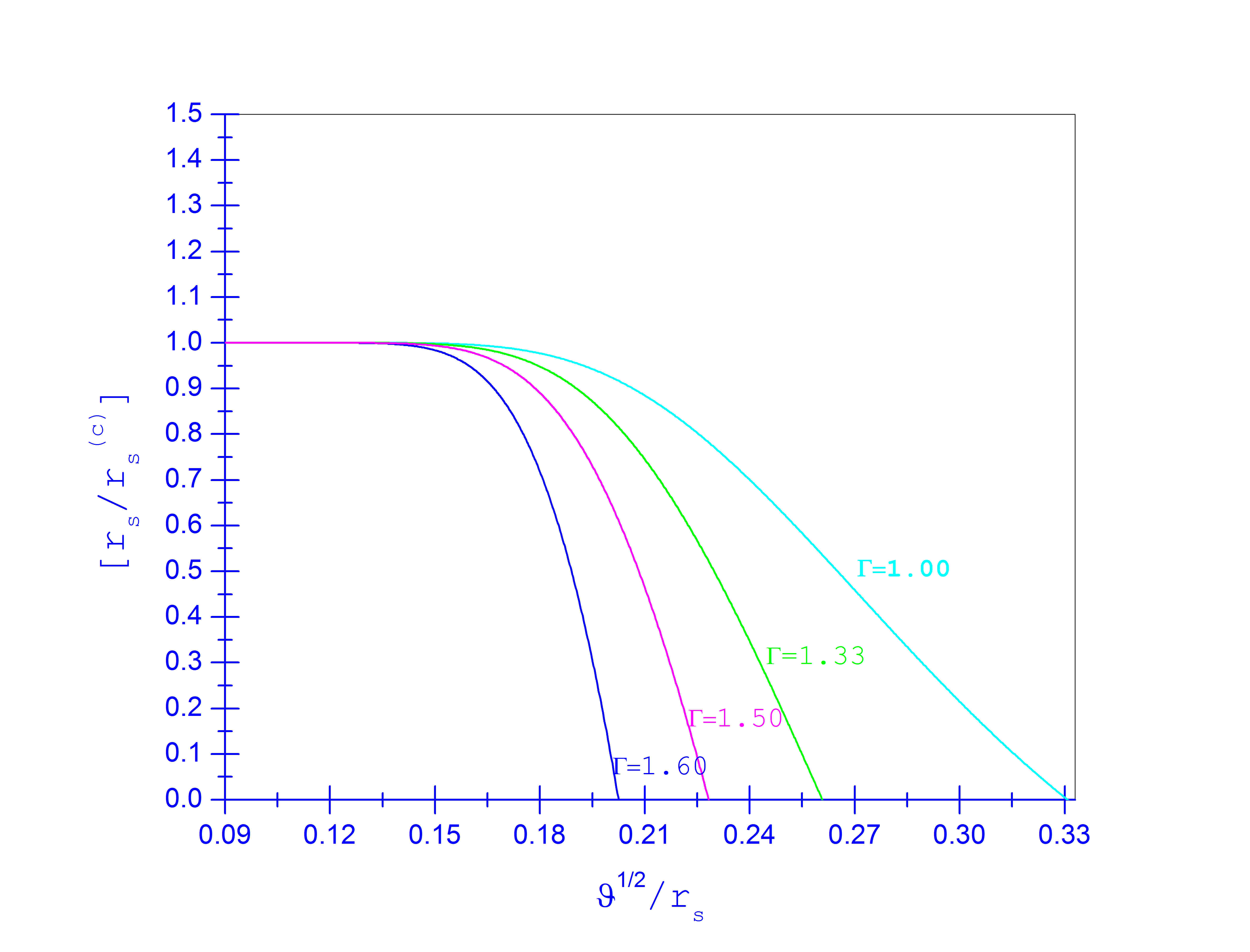}
\includegraphics[scale=.15]{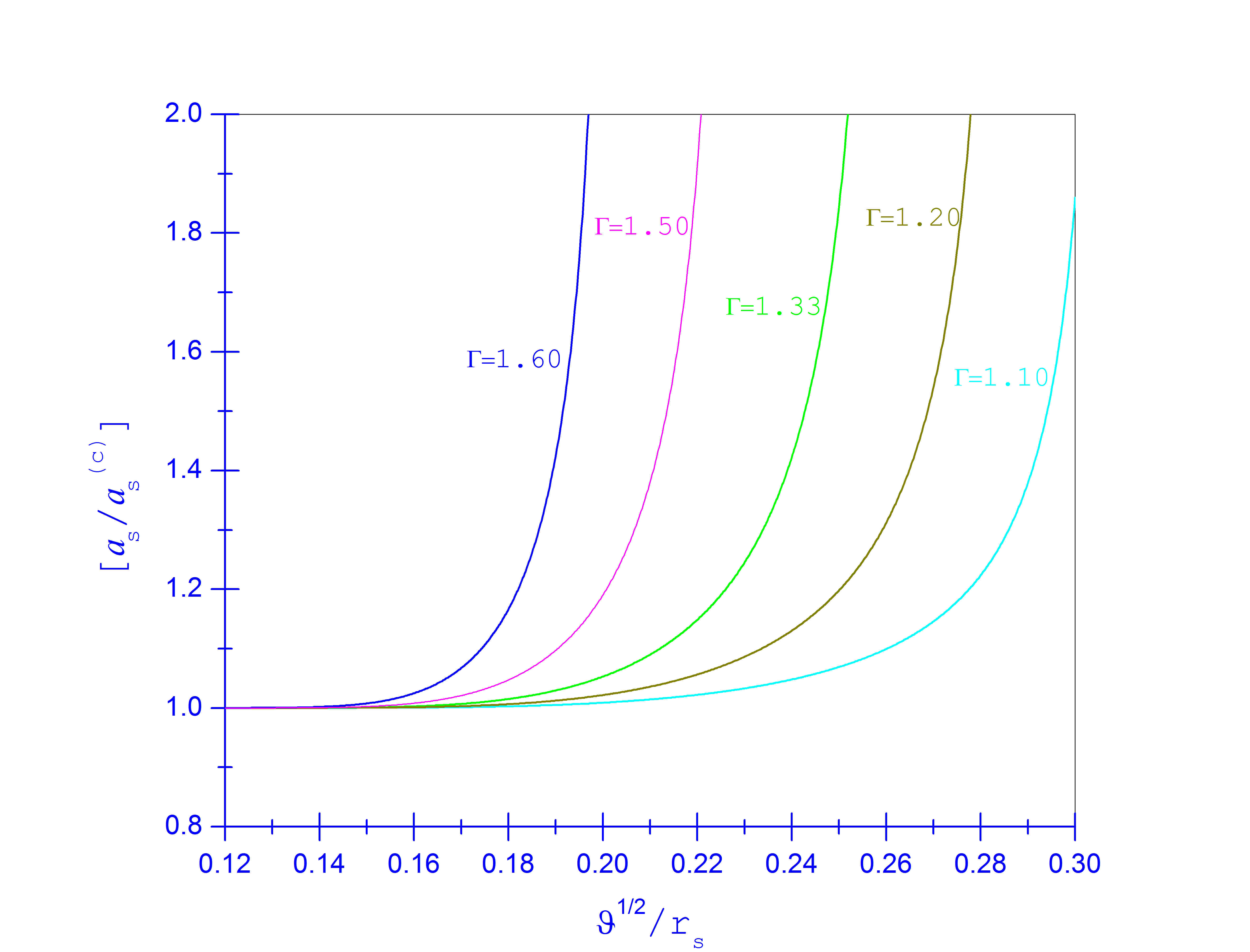}
\caption{\textit{The ratio of the noncommutative sonic radius $r_s$ with the commutative sonic radius $r_s^{(c)}$ decreases with the increase in the  strength of noncommutativity as depicted in the figure at the top. But the ratio of the noncommutative speed of sound $a_s$ with the commutative one $a_s^{(c)}$ at the sonic point $r_s$ is seen to increase with the increase in the strength of noncommutativity as depicted in the figure at the bottom. }}
\label{FIG.1}
\end{figure}

\noindent 
 For large but finite values of $r$, that is for $r \geq r_s$ the baryons will be nonrelativistic. Hence expanding eq.($\ref{BNLI}$) upto leading order in $a_s$ and $a_{\infty}$, we get 
\begin{equation}
\frac{a_s^2\:}{a^2_{\infty}}=\frac{2}{5-3\Gamma} \frac{1-\frac{r_s\gamma'_s}{\gamma_s}}{1-\frac{r_s\gamma'_s}{\gamma_s}\left(\frac{1\,+\,\Gamma}{5-3\Gamma} \right)}.
\label{sonic_spd}
\end{equation} 
The critical (sonic) radius $r_s$ can now be obtained by using eq.($\ref{sonic_spd}$) in eq.($\ref{r_s2}$):
\begin{eqnarray}
r_s &\approx & \frac{5-3\Gamma}{4}\frac{2M\gamma_s\left[1-\left(\frac{1\,+\,\Gamma}{5-3\Gamma} \right)r_s\gamma'_s/\gamma_s\right]}{\sqrt{\pi} a^2_{\infty}}~;~ \Gamma<5/3 \\
&=&\frac{2\gamma_s}{\sqrt{\pi}}\left[1-\left(\frac{1\,+\,\Gamma}{5-3\Gamma} \right)r_s\gamma'_s/\gamma_s\right] r_s^{(c)}
\label{r_sa}
\end{eqnarray} where 
\begin{equation}
r_s^{(c)}=\frac{5-3\Gamma}{4}\frac{M}{a_{\infty}^2}
\end{equation} is the sonic radius in the commutative ($\vartheta=0$) case. The profiles of the sonic radius and the speed of sound at the sonic point are presented in Figure $\ref{FIG.1}$. For $r_s >> \sqrt{\vartheta}$, eq.(s)($\ref{sonic_spd}$) and ($\ref{r_sa}$) reduce to
\begin{equation}
 a_s^2\approx \frac{2}{5-3\Gamma}a^2_{\infty}  ~~;~~ \Gamma< 5/3
\end{equation}
\begin{equation}
 r_s\approx\frac{5-3\Gamma}{4}\frac{M}{ a^2_{\infty}} ~~;~~ \Gamma< 5/3.
\end{equation}
 The next thing is to find the profile of the number density of the adiabatic gas at the sonic point. For $a_s^2/(\Gamma-1)<<1$,  we get from eq.($\ref{n_a}$) 
 \begin{equation}
\quad\frac{n_s\;\:}{n_{\infty}}= \left( \frac{a_s}{a_{\infty}} \right)^{2/(\Gamma-1)}.
\label{n_sn}
\end{equation} With the above results in hand, we are now in a position to compute the rate of accretion of matter. 

\subsection{Mass accretion rate} 
\begin{figure}
\includegraphics[scale=.15]{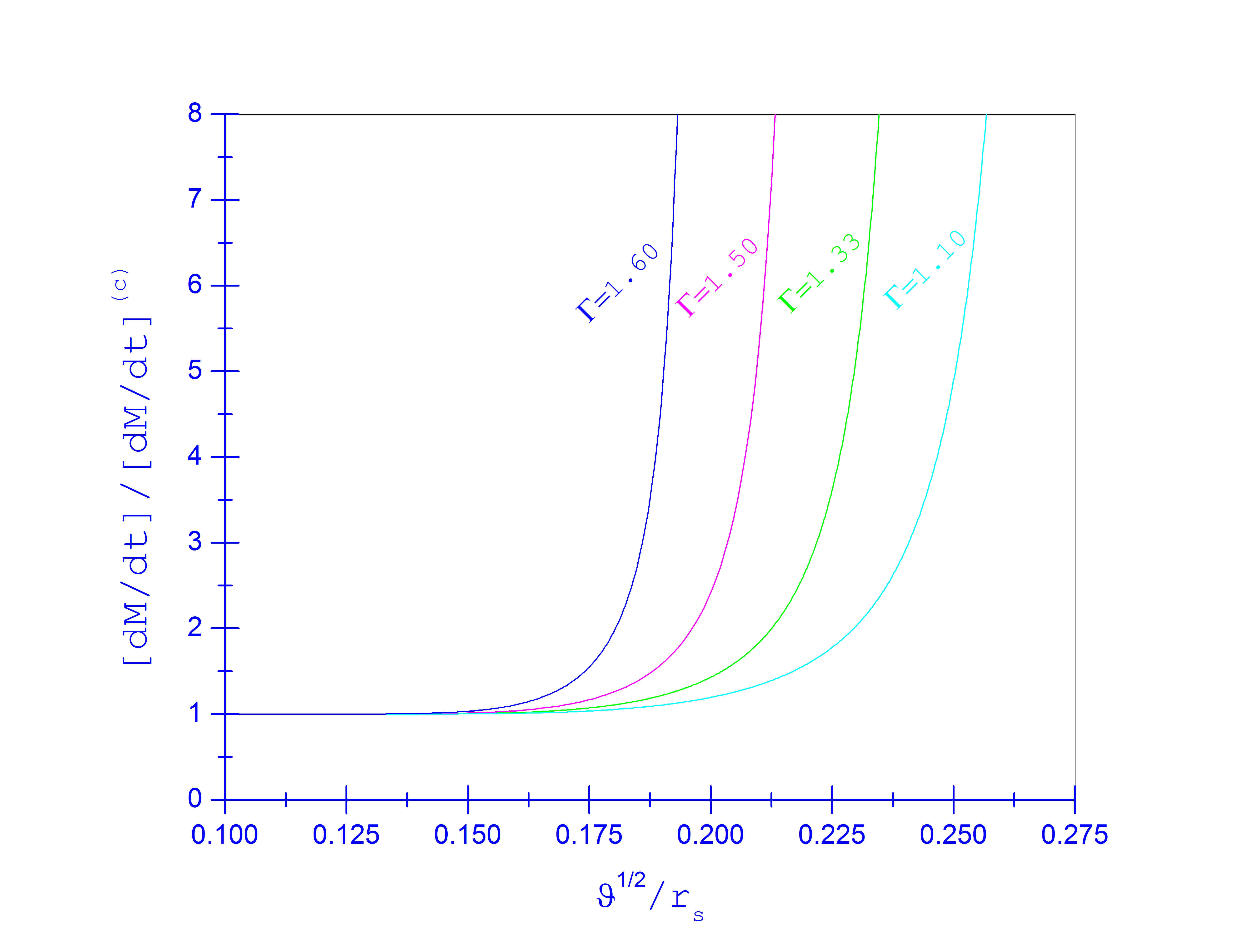}
\caption{\textit{The ratio of the NC matter accretion rate $\dot{M}$ with the commutative mass accretion rate $\dot{M}^{(c)}$ (eq.($\ref{intrrln}$)) rises rapidly as the strength of noncommutative parameter increases. }}
\label{FIG.2}
\end{figure}

The mass accretion rate $\dot{M}$ as given in eq.($\ref{RATE1}$) is independent of $r$ and hence the sonic point $r=r_s$ is used to compute it. We shall also consider two sound speed regimes. In the regime  $a^2_{\infty}<<1$,  the mass accretion rate is given by (with the help of eq.(s)($\ref{v_s2}$, $\ref{r_s1}$, $\ref{RATE1}$, $\ref{n_a}$) )
\begin{eqnarray}
\dot{M} &=& 4\pi r^2_sm_bn_sv_s \\
&=& 4\pi m_bn_{\infty}\left[ \frac{m(r_s)}{2} \right]^2 \phi^2(r_s)[1+a_sf(r_s)]^{-1/2}a_s^{-3} [1-a_s^2/(\Gamma-1)]^{-1/(\Gamma-1)} \left( \frac{a_s}{a_{\infty}} \right)^{2/(\Gamma-1)} 
\label{Gen_acnrate}
\end{eqnarray} where
\begin{equation}
\phi(r_s)=1+3a^2_s-r_s\gamma'_s/\gamma_s(1-a^2_s)
\end{equation}and
\begin{equation}
f(r_s)\equiv \left\lbrace \frac{3+r_s\gamma'_s/\gamma_s}{1-r_s\gamma'_s/\gamma_s} \right\rbrace.
\end{equation} \\ 
\noindent For $a_{\infty}^2<a_s^2<<1$, either from eq.(s)($\ref{Gen_acnrate}$), ($\ref{sonic_spd}$) and ($\ref{r_sa}$) or on directly using eq.(s) ($\ref{v_s2}$), ($\ref{sonic_spd}$), ($\ref{r_sa}$) and ($\ref{n_sn}$) we get  
\begin{eqnarray}
\dot{M} &=& 4\pi r^2_sm_bn_sv_s \\
&=& 4\pi\left(\frac{2\gamma_s}{\sqrt{\pi}} \right)^2\left[1-\left(\frac{1\,+\,\Gamma}{5-3\Gamma} \right)r_s\gamma'_s/\gamma_s\right]^2\eta_s\left(\frac{GM}{a^2_{\infty}}\right)^2\lambda_s m_b n_{\infty} a_{\infty}
\label{RATE2}
\end{eqnarray} where
\begin{equation}
\lambda_s=\left( \frac{1}{2} \right)^{\Gamma+1/2(\Gamma-1)}\left(\frac{5-3\Gamma}{4}\right)^{-(5-3\Gamma)/2(\Gamma-1)}
\label{lambda_s}
\end{equation}
\begin{equation}
\eta_s= \left[\frac{1-\frac{r_s\gamma_s'}{\gamma_s}}{1-\frac{r_s\gamma_s'}{\gamma_s}\left(\frac{1\,+\,\Gamma}{5-3\Gamma} \right)}\right]^{\frac{\Gamma+1}{\Gamma -1}}.
\label{eta_s}
\end{equation}
It turns out that the result involving the effect of noncommutativity differs from the commutative result by a factor. The above relation can be recast as 
\begin{equation}
\dot{M}=\eta_s\left\lbrace\frac{2\gamma(3/2,r^2_s/4\vartheta)}{\sqrt{\pi}} \right\rbrace^2\left[1-\left(\frac{1\,+\,\Gamma}{5-3\Gamma} \right)r_s\gamma_s'/\gamma_s\right]^2\dot{M}^{(c)}
\label{intrrln}
\end{equation} where 
\begin{equation}
\dot{M}^{(c)}=4\pi \left(\frac{GM}{a^2_{\infty}}\right)^2\lambda_s m_b n_{\infty} a_{\infty}.
\label{comm_acnrate}
\end{equation} Let us further introduce
\begin{equation}
\zeta_s=\left\lbrace\frac{2\gamma(3/2,r^2_s/4\vartheta)}{\sqrt{\pi}} \right\rbrace^2\left[1-\left(\frac{1\,+\,\Gamma}{5-3\Gamma} \right)r_s\gamma_s'/\gamma_s\right]^2.
\label{zeta_s}
\end{equation}
 This helps us to rewrite eq.($\ref{intrrln}$) as 
 \begin{equation}
\dot{M}=\eta_s\zeta_s\dot{M}^{(c)}~.
\label{intrrln2}
\end{equation} This relation shows that the mass accretion rate increases with the increase in the strength of noncommutativity and also shows that noncommutativity makes the mass accretion rate depend on the position of the sonic point $r_s$. These features are completely new and should be relevant in the primordial black holes. Reassuringly, we recover the commutative result in the limit $\vartheta=0$ or when $r_s\sim M >>\sqrt{\vartheta}$ . The profile of accretion rate with the varying strength of the  noncommutative parameter is presented in Figure $\ref{FIG.2}$.  

\subsection{Numerical Analysis}
Before we proceed to make numerical computations, we take note of some important points. Our first aim is to solve eq.($\ref{r_sa}$) graphically in order to obtain the numerical values of $r_s$ . 

\noindent
 We shall use $m_b=m_p= 1.67\times 10^{-24}$ gm, $a_{\infty}=10^{7}$ cm/s, $n_{\infty}=10^{30} cm^{-3}$ for different choices of $\Gamma$. The choice of $n_{\infty}$ is particularly crucial in analyzing the matter accretion phenomenon for NC black holes. Writing $\dot{M}= m_b\dot{N}$ implies that $\dot{N}\sim 4\pi r_H^2 v_H\times n\sim 4\pi r_H^2 c\times n$. Corresponding to $r_H\sim \sqrt{\vartheta} \sim 10^{-20} cm$ in order to have atleast $\dot{N}\sim 1 sec^{-1}$, we need $n\sim 10^{29} cm^{-3}$. 
 
 \noindent
 It is easy to note from eq.($\ref{RATE2}$) that $\dot{M}$ will vary for different values of the adiabatic index $\Gamma$. In this paper we are going to carry out computations of $\dot{M}$ obtained with two specific values of $\Gamma$. 
 
\noindent
For $\Gamma=4/3$, we have $\lambda_s=0.71$ from eq.($\ref{lambda_s}$), and eq.($\ref{comm_acnrate}$) gives $\dot{M}^{(c)}\simeq 1.48\times 10^{-6}$ gm/s for $(GM/c^2)=10^{-20} cm$, which is incidentally the value of matter accretion rate onto a commutative black hole $\cite{SHP_TEU, John}$. On the other hand for $\Gamma=1.5$, we have $\lambda_s=0.50$, and eq.($\ref{comm_acnrate}$) gives $\dot{M}^{(c)}\simeq 1.042\times 10^{-6}$ gm/s. The numerical results are presented in Tables $1$ and $2$.
\bigskip \bigskip \bigskip \bigskip \bigskip \bigskip \bigskip \bigskip \bigskip \bigskip \bigskip \bigskip \bigskip \bigskip \bigskip \bigskip \bigskip \bigskip \bigskip

\begin{center}
\noindent Table I:  $\Gamma=4/3$ , $\lambda_s=0.71$ , $\dot{M}^{(c)}= 1.48\times 10^{-6}$ gm/s \bigskip

\begin{tabular}{|c|c|c|c|c|}
\hline
 $ \sqrt{\vartheta}/r_s$ &$ r_s/\sqrt{\vartheta}$ &$ \eta_s$ & $\zeta_s$&   $\dot{M} \;in \;\mathrm{gm/s}$  \\
 \hline 
$0.000 $ & $\infty $ & $ 1.000 $ & $ 1$& $1\times1\times1.48\times 10^{-6}= 1.48\times 10^{-6} $ 

\\
\hline 
$0.100 $ & $10 $ & $ 1.000 $ & $ 1$& $1\times1\times1.48\times 10^{-6}= 1.48\times 10^{-6} $ 

\\

\hline $ 0.140$ & $ 7.1429$ & $1.003 $ & $ 0.9986 $ & $1.003\times0.9986\times1.48\times 10^{-6}=1.4824\times 10^{-6} $
\\
\hline 
$ 0.150$ & $6.6667$& $ 1.011$ & $ 0.994$ & $1.011\times0.994\times  1.48\times 10^{-6}=1.4873\times 10^{-6} $ 
\\
\hline
$ 0.160$ & $6.2500 $& $ 1.037$ & $0.981 $ & $1.037\times0.981\times 1.48\times 10^{-6}= 1.5056\times 10^{-6} $ 
\\
\hline
$ 0.170$ & $5.8824 $& $ 1.100$ & $0.953 $ & $1.100\times0.953\times 1.48\times 10^{-6}=1.5515\times 10^{-6} $ 
\\
\hline $ 0.180$ & $ 5.5556$ & $ 1.231$ & $ 0.900$ & $1.231\times0.900\times 1.48\times 10^{-6}=1.6400\times 10^{-6} $ 
\\
\hline
$ 0.190$ & $5.2632 $ & $ 1.50$ & $0.815 $ & $1.50\times0.815\times 1.48\times 10^{-6} = 1.8093\times 10^{-6}$ 
\\
\hline $ 0.200$ & $ 5.0000$ & $ 2.04$ & $0.697 $ & $2.04\times0.697\times 1.48\times 10^{-6}=2.104\times 10^{-6} $ 
\\
\hline
$ 0.210$ & $ 4.7619$& $ 3.3$ & $ 0.554$ & $3.3\times0.554\times 1.48\times 10^{-6}=2.71\times 10^{-6} $ 
\\\hline 
$ 0.220$ & $4.5454 $& $ 6.8$ & $0.398 $& $6.8\times0.398\times 1.48\times 10^{-6}= 4.01\times 10^{-6} $ 
\\\hline
$ 0.225$ & $4.4444 $ & $ 11.2$ & $0.320 $ & $11.2\times0.320\times 1.48\times 10^{-6}=5.3\times 10^{-6} $ 
\\
\hline 
$ 0.230$ & $4.3478 $ & $ 20.6$ & $0.247 $ & $20.6\times0.247\times 1.48\times 10^{-6}=7.53\times 10^{-6} $ 
\\
\hline 
$ 0.233$ & $4.2918 $& $32.4$ & $ 0.205$ & $32.4\times0.205\times 1.48\times 10^{-6}=9.83\times 10^{-6} $ 
\\
\hline
$ 0.235$ & $4.2553 $& $ 45.5$ & $ 0.179$& $45.5\times0.179\times 1.48\times 10^{-6}=12.1\times 10^{-6} $ 
\\
\hline
$ 0.237$ & $ 4.2194$ & $ 65.9$ & $0.154 $& $65.9\times0.154\times 1.48\times 10^{-6}=15.0\times 10^{-6} $ 
\\
\hline
$ 0.240$ & $4.1667 $& $ 127$ & $0.120 $& $127\times0.120\times 1.48\times 10^{-6}=22.6\times 10^{-6} $ 
\\
\hline 
$ 0.245$ & $4.0816 $& $ 522$ & $ 0.070$& $522\times0.070\times 1.48\times 10^{-6}=54.1\times 10^{-6} $ 
\\
\hline
\end{tabular}

\end{center}

\bigskip

\begin{center}
\noindent Table II: For adiabatic index $\Gamma=1.50$,  $\lambda_s=0.50$ , $\dot{M}^{(c)}= 1.042\times 10^{-6} gm/s$
\end{center}
\begin{center}
\begin{tabular}{|c|c|c|c|c|}
\hline
 $ \sqrt{\vartheta}/r_s$ & $r_s/\vartheta $ &$ \eta_s$ & $\zeta_s $ &   $\dot{M} \;in \;\mathrm{gm/s}$  \\
 \hline 
$0.000 $ & $ \infty $& $ 1.000 $ & $1.000 $ & $1\times1\times1.042\times 10^{-6} = 1.042\times 10^{-6}$ 

\\
\hline 
$0.100 $ & $ 10$& $ 1.000 $ & $1.000 $ & $1\times1\times1.042\times 10^{-6} = 1.042\times 10^{-6}$ 

\\
\hline
$ 0.140$ & $ 7.1429$ & $ 1.006$ & $ 0.997$ & $1.006\times0.997\times 1.042\times 10^{-6}=1.045\times 10^{-6} $ 
\\

\hline
 $ 0.150$ & $6.6667 $ & $1.025 $ & $ 0.987$ & $1.025\times0.987\times1.042\times 10^{-6}=1.054\times 10^{-6} $
\\
\hline 
$ 0.160$ & $ 6.25$ & $ 1.082 $ & $0.960 $ & $1.082\times0.960\times  1.042\times 10^{-6}=1.082\times 10^{-6} $ 
\\
\hline
$ 0.170 $ & $ 5.8824$ & $ 1.232$ & $0.900$ & $ 1.232\times0.900\times 1.042\times 10^{-6}=1.234\times 10^{-6} $ 
\\
\hline
$ 0.180$ & $5.5556 $ & $ 1.585$ & $0.794 $ & $1.58\times0.794\times 1.042\times 10^{-6}=1.31\times 10^{-6} $ 
\\
\hline
$ 0.185$ & $5.4054 $&  $ 1.94$ & $0.718 $ & $1.94\times0.718\times 1.042\times 10^{-6}=1.451\times 10^{-6} $ 
\\
\hline $ 0.190$ & $5.2632 $ & $ 2.54$ & $0.632 $ & $2.54\times0.632\times 1.042\times 10^{-6}=1.673\times 10^{-6} $ 
\\
\hline
$ 0.195$ & $ 5.1282$ & $ 3.58$ & $ 0.534$ & $3.58\times0.534\times 1.042\times 10^{-6}=1.95\times 10^{-6} $ 
\\
\hline
$ 0.200$ & $5.0000 $& $ 5.7$ & $0.428 $ & $5.7\times0.428\times 1.042\times 10^{-6}=2.44\times 10^{-6} $ 
\\
\hline
$ 0.205$ & $4.8780 $ & $ 10.7$ & $0.319 $ & $10.7\times0.319\times 1.042\times 10^{-6} =3.56\times 10^{-6}$ 
\\
\hline $ 0.210$ & $ 4.7619$ & $ 25.2 $ & $0.216 $& $25.2\times0.216\times 1.042\times 10^{-6}=5.7\times 10^{-6} $ 
\\
\hline
$ 0.213$ & $4.6948 $& $ 50.1$ & $0.159 $ & $50.1\times0.159\times 1.042\times 10^{-6}=8.3\times 10^{-6} $ 
\\
\hline 
$ 0.215$ & $ 4.6512$ & $ 87.9 $ & $0.123 $ & $87.9\times0.123\times 1.042\times 10^{-6}=11.3\times 10^{-6} $ 
\\\hline 
$ 0.217$ & $ 4.6083$ &$ 172$ & $ 0.092$ & $172\times0.092\times 1.042\times 10^{-6}=16.5\times 10^{-6} $ 
\\
\hline
$ 0.220$ & $4.5454 $ & $ 658 $ & $ 0.052$& $658\times0.052\times 1.042\times 10^{-6}=35.7\times 10^{-6} $ 
\\
\hline

\end{tabular}
\end{center}

\section{Gas behaviour around the sub Bondi radius ($r_H<r_{sb}<r_s$)} 
\begin{figure}
\includegraphics[scale=.15]{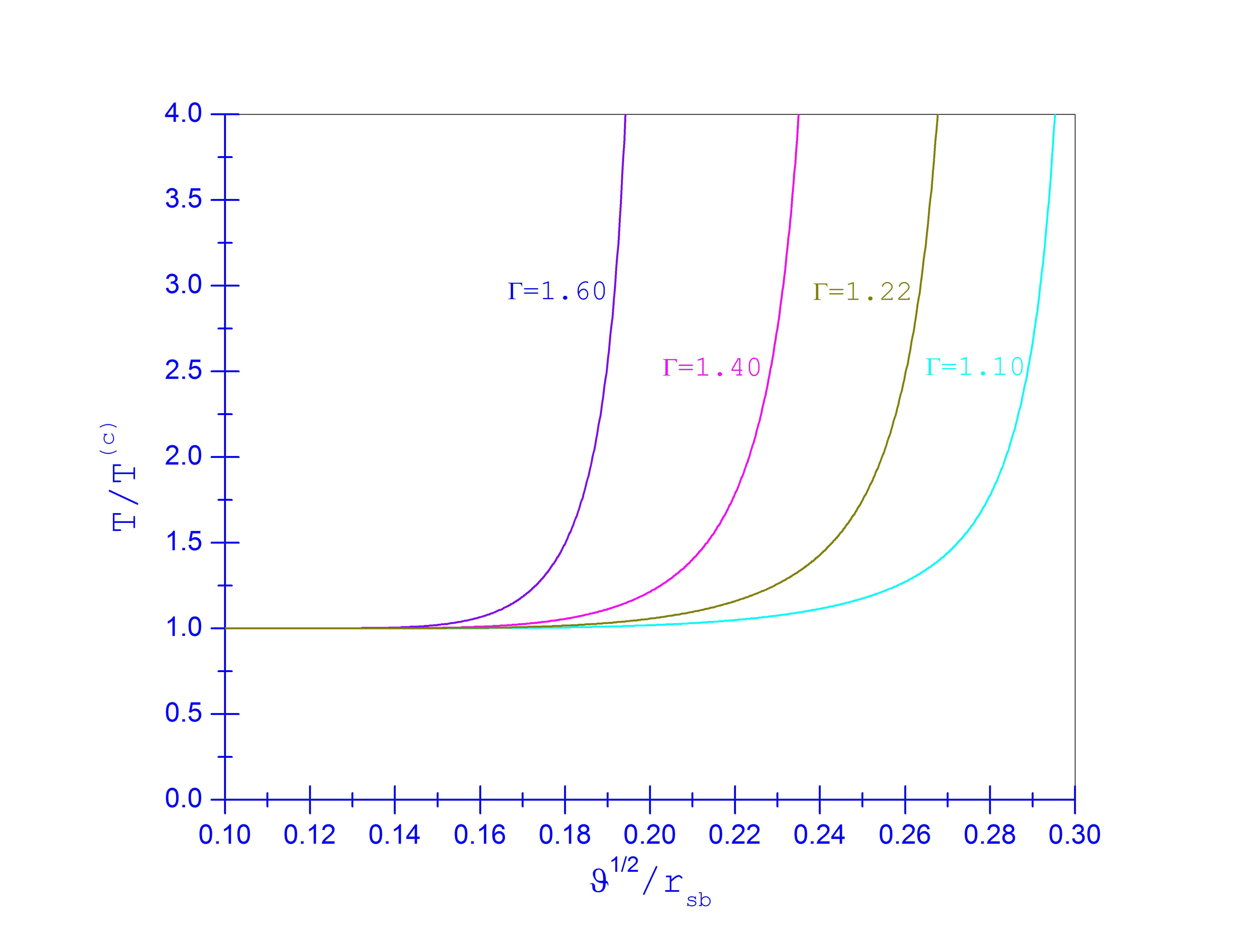}
\caption{\textit{The temperature at a sub-Bondi radius increases quite rapidly with the increasing noncommutativity. Different lines correspond to different values of $\Gamma$.}}
\label{FIG_3}
\end{figure}

In this section we estimate the flow characteristics in the range $r_H < r << r_s$. At distances less than the Bondi radius $r_s$, the accreting gas is supersonic. The transonic flow has a radial velocity more than the speed of sound, that is $v>a$. The upper bound on the radial dependence of gas velocity can be estimated using eq.($\ref{Dummy_Brnlli}$). Since by standard $a_{\infty}^2<a_{sb}^2<<1$, eq.($\ref{Dummy_Brnlli}$) implies 
\begin{equation}
\left(1-\frac{2m(r)}{r}+v^2  \right)<1.
\end{equation}  Thus the upper bound to the transonic flow is determined to be
\begin{equation}
v^2\approx \frac{2m(r)}{r} ~.
\label{v_H}
\end{equation} By equating the expressions of $\dot{M}$ in eq.($\ref{RATE1}$) and ($\ref{RATE2}$) and using eq.($\ref{v_H}$), we get
\begin{equation}
\frac{n(r_{sb})}{n_{\infty}}\approx \frac{\lambda_s\eta_s}{\sqrt{2}}\left( \frac{M}{a^2_{\infty}r_{sb}} \right)^{3/2}\left[1-\left(\frac{1\,+\,\Gamma}{5-3\Gamma} \right)r_{sb}\gamma'_s/\gamma_{sb}\right]^2\left(\frac{2\gamma_{sb}}{\sqrt{\pi}} \right)^{3/2}.
\label{rel_density}
\end{equation} The above equation reduces to \[\frac{n(r_{sb})}{n_{\infty}}\approx \frac{\lambda_s}{\sqrt{2}}\left( \frac{M}{a^2_{\infty}r_{sb}} \right)^{3/2}  \] in the $r_{sb}>>\sqrt{\vartheta}$ limit $\cite{SHP_TEU}$. For a Maxwell-Boltzmann gas, we have a gas state equation given by $p=nk_B T$. Recalling eq.($\ref{p_n}$), yields the adiabatic temperature profile to be 
\begin{eqnarray}
\frac{T(r_{sb})}{T_{\infty}} &\approx & \left( \frac{n(r_{sb})}{n_{\infty}} \right)^{\Gamma-1}\\
&=& \left(\frac{2\gamma_{sb}}{\sqrt{\pi}} \right)^{3(\Gamma-1)/2}\left( \frac{M}{a^2_{\infty}r_{sb}} \right)^{3(\Gamma-1)/2}\left[1-\left(\frac{1\,+\,\Gamma}{5-3\Gamma} \right)r_s\gamma'_s/\gamma_s\right]^{2(\Gamma-1)}\left(\frac{\lambda_{s}\eta_s}{\sqrt{2}} \right)^{\Gamma-1}.
\end{eqnarray} 
The temperature profile is shown in Figure $\ref{FIG_3}$. Evidently, this result also reduces to the commutative result when $r_{sb}>>\sqrt{\vartheta}$  $\cite{SHP_TEU}$. 

\section{Gas behavior at the event horizon} In this section we estimate the flow characteristics at the event horizon $r_H$. The horizon of the NC inspired Schwarzschild black hole corresponds to $r_{H}=2m(r_H)$. Thus the flow velocity $v^2\approx 2m(r)/r$ approaches the speed of light right at the event horizon. Hence, using eq.(s) ($\ref{RATE1}$, $\ref{RATE2}$, $\ref{v_H}$), we get 
 \begin{equation} 
\frac{n(r_H)}{n_{\infty}}\approx \frac{\lambda_H\eta_H}{4}\left( \frac{c}{a_{\infty}}  \right)^3\left[1-\left(\frac{1\,+\,\Gamma}{5-3\Gamma} \right)r_H\gamma'_H/\gamma_H\right]^2
\end{equation}
 \begin{equation}
\frac{T(r_H)}{T_{\infty}}\approx \left(\frac{\lambda_s\eta_s}{4}\right)^{(\Gamma-1)}\left( \frac{c}{a_{\infty}}  \right)^{3(\Gamma-1)}\left[1-\left(\frac{1\,+\,\Gamma}{5-3\Gamma} \right)r_H\gamma'_H/\gamma_H\right]^{2(\Gamma-1)}.
\end{equation} 
Thus we find that noncommutativity also affects the number density and temperature profiles of the accreting gas at the horizon and both the gas temperature and the gas density are larger in magnitude compared to the commutative results.

\section{Conclusions} 
There is a high chance that mini black holes would be produced inside the ultra-high energy particle accelerators, for example in the large hadron collider. These black holes are distinct because of their unique ability of producing remarkable gravitational effects with such an incredibly tiny size. If these tiny black holes really get produced inside the large hadron collider, it is quite plausible that they would accrete matter towards themselves and would create a dense hot environment of gas radiation powered by the accretion process. Even though this thermal environment would survive for a very short duration of time, its analysis is definitely a worthwhile exercise to carry out. Indeed such an analysis may provide a way of testing the effects of gravity at very  small scales of length. 

 Our studies in this paper fulfill our expectations. The noncommutative structure of spacetime have prominent effects on the phenomenon of accretion of matter. The first important observation that we make is that the noncommutativity of spacetime results in a rapid fall in the value of the sonic radius. On the other hand the speed of sound at the sonic-point is found to rise  steeply due to the presence of noncommutativity. The mass accretion rate corresponding to the noncommutative inspired Schwarzschild black hole gets modified to $\mathrm{\dot{M}=\zeta_s\eta_s\dot{M}^{(c)}}$, where the factors $\zeta_s$ and $\eta_s$ capture the effects of a decreasing sonic radius and that of a steeply rising speed of sound. In particular the mass accretion rate is found to increase with the increase in the strength of noncommutativity. Finally the analysis of the thermodynamic profiles of adiabatic gas-matter shows that both the gas density and the gas temperature get markedly enhanced in the presence of noncommutativity. 
 
 \bigskip
 \section*{Acknowledgment} S.G. acknowledges the support by DST SERB under Start Up Research Grant (Young Scientist), File No. YSS/2014/000180.

\end{document}